\documentclass[useAMS,usenatbib]{mn2e}
\usepackage{multirow}
\usepackage{graphicx}
\usepackage{footnote}
\usepackage{hyperref}
\usepackage{amsmath}
\usepackage{amssymb}
\usepackage{soul}
\usepackage{caption}
\usepackage{subcaption}
\usepackage[usenames,dvipsnames]{pstricks}
\graphicspath{{Images/}}
\newcommand{\bd}[1]{\mbox{\boldmath $#1$}}
\newcommand{\RR}[1]{\mathbb{R}^{#1}} 

\title{An Explorative Approach for Inspecting Kepler Data}
\author{S.~D. K\"ugler, N. Gianniotis, K.~L. Polsterer\\
Heidelberg Institute for Theoretical Studies, Schloss-Wolfsbrunnenweg 35,
69118 Heidelberg, Germany}
\begin{document}

\date{}

\pagerange{\pageref{firstpage}--\pageref{lastpage}} \pubyear{2015}

\maketitle

\label{firstpage}

\begin{abstract}
The Kepler survey has provided a wealth of astrophysical knowledge by
continuously monitoring over 150,000 stars. The resulting database contains
thousands of examples of known variability types and at least as many that 
cannot be classified yet. In order to reveal the knowledge hidden in the
database, we introduce a new visualisation method that allows us to inspect
regularly sampled time series in an explorative fashion. To that end, we propose
dimensionality reduction on the parameters of a model capable of representing
time series as fixed-length vector representation. We show that a more refined
objective function can be chosen by minimising the
reconstruction error, that is the deviation between
prediction and observation, of the observed time
series instead of reconstructing model parameters.
The proposed visualisation exhibits a strong correlation between the
variability behaviour of the light curves and their physical properties.
As a consequence, temperature and surface gravity can, for some stars, be
directly inferred from non- (or quasi-) periodic light curves.
\end{abstract}


\section{Introduction}

The launch of the Corot and Kepler spacecrafts (\citealp{2009A&A...506..411A}
and \citealp{2010Sci...327..977B}) initiated a new era in the study of
stellar variability. The unobstructed view to the light of astronomical
sources enabled the continuous monitoring of stellar sources with short cadence
and  very low photometric error. While the primary goal of both
missions was the detection of solid exoplanets, the continuous monitoring of
$150,000$ variable main sequence stars allowed a very detailed study of their
variability behaviour. While astroseismology
\citep{2010PASP..122..131G} and exoplanet detection
\citep{2012ApJ...750..112L} greatly benefit from those observations, many
objects remain unlabelled. The labelling of (quasi-) periodic sources is quite reliable
\citep[e.g.,][]{2010MNRAS.409.1585B,2011AJ....142..160S}, however, a large
fraction of the objects that show no periodic behaviour remains unclassified.

In order to investigate the nature of the objects that cannot be explained by
known variability mechanisms, alternative approaches have to
be found. Visualisation (i.e. dimensionality reduction) and clustering are the
most prominent representative algorithms from the camp of unsupervised learning.
Visualisation in particular allows an intuitive inspection of the properties of
the observed data by projecting them in a lower dimensional space
\citep{DBLP:journals/eswa/KramerGP13}. Data
analysts can interpret the visualisation plot and look for structures and
similarities that could not be detected in the original data space. However, it
is not possible to directly apply dimensionality reduction to  raw
sequential data as the individual measurements are not independent and
therefore the time series cannot be treated like vectorial data. To circumvent
this problem, \citet{2012AJ....143..123M} apply local linear embedding to light
curves of binary stars that have been previously phase-folded and pre-aligned.
Alignment is only possible if the light curve is periodic and a uniquely
identifiable point exists (e.g. point of deepest eclipse for binaries), with
respect to which each light curve of a given class can be aligned.
However, besides phase-folding and alignment, the main issue still persists
namely that the sequential nature of the data has been ignored. This implies
that the dynamical behaviour of the physical systems will not be (adequately)
captured in the visualisation plots.

In this work, we use the echo state network (ESN, \citealp{Lukosevicius2009}) to
describe time series as sequences. The ESN is a discrete time recurrent neural
network that is used to forecast the behaviour of time series.
The ESN, as other neural networks, is parametrised by a vector of weights.
Training the ESN on a time-series yields an optimised vector of weights which
in this work we use as a fixed-length vector representation for regularly
sampled time series.
The advantage of this new representation is that it is invariant to the
variable length of time series as well as to the presence of time shifts.
Instead of performing the visualisation on the original data items, we perform
visualisation on this new representation.
This amounts to coupling the visualisation algorithm to the ESN model and
thereby obtaining a more meaningful visualisation that does take into account
the time behaviour of the time series data. With this tool at hand, we
visualise a dataset of light curves from the Kepler survey and highlight the
meaning of this visualisation in the context of the physical properties of these stars.

In \mbox{Section \ref{sec:meth}}, the proposed methodology is explained in more
detail, while \mbox{Section \ref{sec:data}} describes the Kepler survey.
The results of the visualisation are presented in \mbox{Section
\ref{sec:results}} followed by a discussion with respect to  physical
properties in \mbox{Section \ref{sec:discussion}}. Finally, the prospects of
the presented methodology in the analysis of time series and other astronomical
data are discussed in \mbox{Section \ref{sec:conclusions}}

\section{Methodology}
\label{sec:meth}

In the following, we provide a brief description of the ESN-based visualisation
method.
A more detailed account of the adopted methodology in this work has previously appeared  in
\citet{auto_enc}. The terms time-series and sequence are used interchangeably.

\subsection{Encoding sequences as readouts}

It is evident that dimensionality reduction designed for vectorial data cannot
be directly applied on time series as this would not correctly take into account their sequential nature. 
Therefore, prior to dimensionality reduction, an appropriate vectorial representation of time-series needs to be found that is able to deal with variable lengths and shifts along the time
axis. To that purpose, we employ the ESN architecture \citep{Lukosevicius2009}. 
An ESN is a discrete time recurrent neural network that learns to predict on
regularly sampled time series: given an observation $y_t$ at time $t$, it
makes a prediction for $y_{t+1}$. ESNs  have the great advantage that,  in contrast to other neural network architectures, the hidden non-linear part (known as reservoir) is fixed and only the output weights need to be trained. The output weights,  the so called readouts, interact linearly with the reservoir and are the only free parameters in the model.
Hence, the ESN may be viewed as a linear model written as  $y_{t+1} = \bd{\phi}(y_t) \bd{w}$, where $\bd{w}\in\RR{h\times 1}$ are the readouts and $\bd{\phi}(y_t)\in\RR{1\times h}$ are the activations of the hidden non-linear part induced by input $y_t$. 
\bd{\phi} encodes various aspects of the neural network architecture of the ESN (number of neurons $h$, connectivity, weight structure, etc.) which we do not discuss here; instead we refer the interested reader to \citet{auto_enc} and references therein.

We use the ESN to encode regularly sampled time series as a fixed-length
vector representation. We use the notation $\bd{y}$ to denote an entire light
curve $\bd{y}=(y_1,\dots,y_T)\in\RR{T\times 1}$ composed of individual observations $y_t\in \RR{}$.
Also, we use $\bd{\Phi}\in \RR{T \times h}$ as the matrix that accumulates row-wise all
activations $\bd{\phi}(y_t)$.
Hence, given a sequence $\bd{y}$, the optimised readouts are found by solving
the least squares problem $\| \bd{y} - \bd{\Phi}\bd{w} \|^2$. The obtained
readout $\bd{w}$ is taken as the new representation for a light curve $\bd{y}$.
Given a dataset of $N$ number of light curves $\{\bd{y}^{(1)},\dots,\bd{y}^{(N)}\}$ the first step in our approach is to encode each light curve as a readout vector.
This results to a new dataset of $N$ readout vectors $\{\bd{w}^{(1)},\dots,\bd{w}^{(N)}\}$.


\subsection{Dimensionality reduction}


Dimensionality reduction algorithms seek to find low-dimensional
representations $\bd{x}\in\RR{q}$ of high-dimensional data items
$\bd{w}\in\RR{d}$, where $q<d$, so that they can be visually inspected, i.e.
$q=2$ or $q=3$. In this case, $\bd{w}$ stands for the readout weights returned
by the ESN. A typical criterion that drives the training of visualisation
algorithms is the reconstruction error: one attempts to reconstruct from the
low-dimensional representations $\bd{x}$ the original vector $\bd{w}$. Of
course, due to the  information loss incurred during the dimensionality
reduction, the reconstruction is only approximate and we obtain reconstructed
readouts $\tilde{\bd{w}}$. A typical choice for the reconstruction error is the
squared $L_2$ norm, $\|\bd{w} - \tilde{\bd{w}}\|^2$.

For instance, principal component analysis (PCA, \citealp{PCA}) projects
$\bd{w}$ linearly to a low-dimensional space spanned by the top two
eigenvectors $\bd{u}_1$,$\bd{u}_2$ of the sample covariance matrix. In this
case, the linear projection is given by matrix $\bd{U}=[\bd{u}_1\
\bd{u}_2]\in\RR{d\times 2}$, the lower dimensional representation reads
$\bd{x}= (\bd{w}\bd{U})^T $, and the reconstruction reads $\tilde{\bd{w}}  =
(\bd{Ux})^T$. Hence, the reconstruction error of PCA is given by $\| \bd{w} -
\bd{U} \bd{U}^T \bd{w}\|^2$.

Another dimensionality reduction algorithm, often viewed as a kind of
non-linear PCA, is the autoencoder \citep{Kramer}.
The autoencoder may be viewed as the composition of two functions, the
projection to low-dimensions denoted by $f_{enc}(\bd{w})=\bd{x}$ (the
non-linear analogue of $\bd{U}$), and the reconstruction to the
high-dimensional space denoted by $f_{dec}(\bd{x})=\tilde{\bd{w}}$ (the
non-linear analogue of $\bd{U}^T$). Hence, the complete mapping is a function
$f(\bd{w};\bd{\theta})=f_{dec}( f_{enc}(\bd{w}))= \tilde{\bd{w}}$, where
$\bd{\theta}$ are the free parameters, the weights of the autoencoder. The
autoencoder's reconstruction error is given by $\|\bd{w} -
f(\bd{w};\bd{\theta})\|^2 \ $ which is the analogue to the PCA objective $\|
\bd{w} - \bd{U} \bd{U}^T \bd{w}\|^2$.

\subsection{ESN-coupled autoencoder}

\begin{figure}
\includegraphics[width=.47\textwidth]{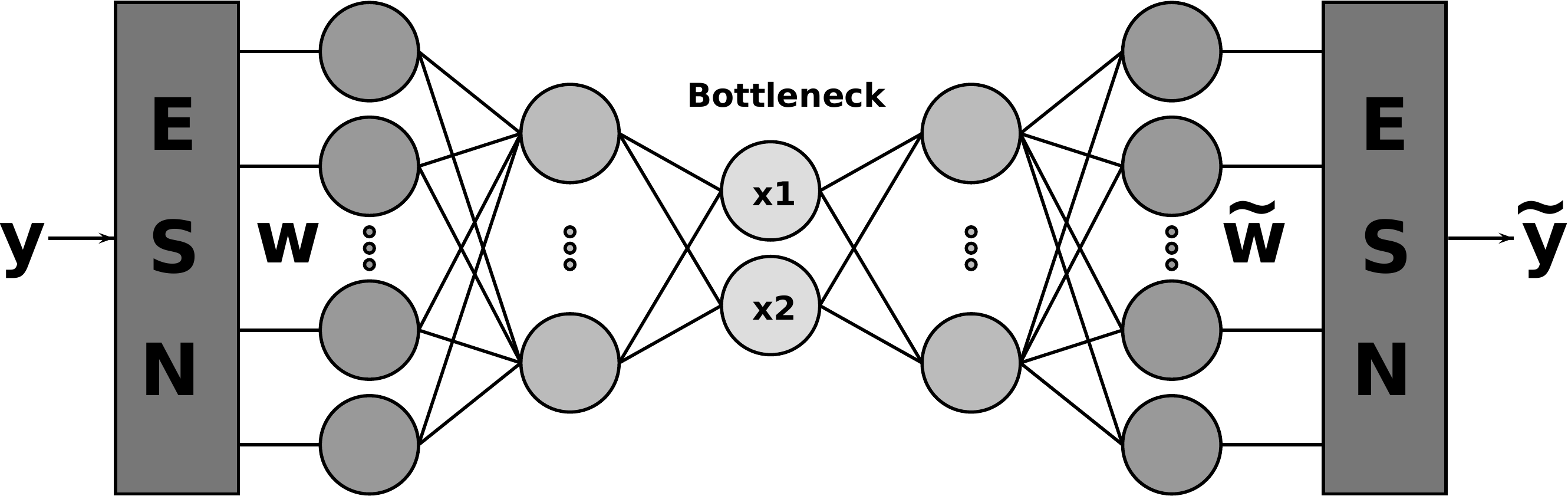}
\caption{Schematic plot of the ESN-coupled autoencoder.\label{fig:AE}}
\end{figure}

So far, the presented algorithms measure how good the reconstruction between the
original weight $\bd{w}$ and the reconstructed weights
$\tilde{\bd{w}}$ is in terms of their $L_2$ distance. A better measure would be
to check how well the reconstruction $\tilde{\bd{w}}$ still represents its
corresponding sequence $\bd{y}$. This can be simply checked by looking at the
reconstruction error $\| \bd{y} - \bd{\Phi}\tilde{\bd{w}} \|^2$ measured with
respect to the sequence $\bd{y}$, as proposed in \citet{auto_enc}.
This measure is much more meaningful as it measures reconstruction in terms of
the prediction quality of the underlying ESN model.
This suggests modifying the reconstruction error of the autoencoder  so that it
now reads:
\begin{equation*}
 \|  \bd{y} - \bd{\Phi} f(\bd{w};\bd{\theta})  \|^2  =  \|  \bd{y} -
 \bd{\Phi}\tilde{\bd{w}} \|^2  =  \|  \bd{y} - \tilde{\bd{y}} \|^2
\end{equation*}

Essentially, the above reconstruction error entails an autoencoder that is
coupled to an ESN as illustrated in \mbox{Fig. \ref{fig:AE}}:  the echo state
network is used to encode light curves \bd{y} as readouts $\bd{w}$.
Next, the readouts are compressed to a low dimensional representation \bd{x}.
Out of the low dimensional representation we then reconstruct a readout 
$\tilde{\bd{w}} $ which when plugged back to the ESN should give a
reconstruction $\tilde{\bd{y}} $ for the input sequence $\bd{y}$.
Optimisation of the autoencoder follows by gradient optimisation via the
backpropagation algorithm \citep{Bishop1996} as typically done for
neural networks. We term the proposed visualisation approach as the ESN-coupled
autoencoder which we abbreviate here as ESN-AE.

\subsection{Summary of employed models}

We briefly summarise the  concepts presented in the
preceding sections. The basic idea of the presented approach is to find a suitable vector representation of the
time series data. Therefore, an ESN is trained on each light curve $\bd{y}$ and the resulting readout $\bd{w}$ is used as a fixed-length vector representation. 

The standard way
of reducing the dimensionality of  vectors is to find  representations
 that  minimise the $L_2$ norm between the original vectors and
their reconstructions. This can be  done either in a linear
(PCA) or non-linear (autoencoder) fashion. However, minimising $\|\bd{w} -
\tilde{\bd{w}}\|^2$ is problematic as it does not capture in any way how well the reconstructed readout $\tilde{\bd{w}}$ 
still represents the corresponding light curve \bd{y}.  
On the other hand, in the proposed ESN-AE model we put forward an
alternative reconstruction that measures how well a reconstructed
readout weight $\tilde{\bd{w}}$ can still predict the original light curves
$\bd{y}$ it represents. In order to distinguish the ESN-AE model from the
autoencoder, we henceforth call the later one ``plain autoencoder''.
 We summarise  these concepts  in \mbox{Tab. \ref{tab:sum}}.
\begin{table}
\centering
\begin{tabular}{l|l|l}
Visualisation & Linear & Objective \tabularnewline
\hline
PCA & Yes &  $\|\bd{w} - \tilde{\bd{w}}\|^2=\| \bd{w} - \bd{U} \bd{U}^T \bd{w}
\|^2$ \tabularnewline 
Plain autoencoder & No & $\|\bd{w} - \tilde{\bd{w}}\|^2=\| \bd{w}
- f(\bd{w};\bd{\theta}) \|^2 \ $ \tabularnewline 
ESN-AE & No & $\|  \bd{y}\,
-\,\, \tilde{\bd{y}} \|^2 = \| \bd{y}\, - \bd{\Phi}f(\bd{w};\bd{\theta})\|^2$
\tabularnewline
\end{tabular}
\caption{Summary of  proposed visualisation algorithms.\label{tab:sum}}
\end{table}

\section{Data}
\label{sec:data}

The data used in this work come from the Kepler satellite mission
described in detail in \citet{2010Sci...327..977B}.
The Kepler mission aims for the detection of Earth-like planets and therefore
continuously monitors $150,000$ main sequence stars in a fixed $115$
square degrees field of view located between Cygnus and Lyrae. Besides the
detection of exoplanets, the mission opened the window to a detailed
study of the dynamical behaviour of variable stars (e.g., RR Lyrae stars). 
Sources were
observed with different cadences over several quarters. In this work, we focus
only on data from the first quarter with long exposures of 29.4\,min each
(33.16 days in total).
The available data volume, even in the chosen subset, is quite large as the first
quarter comprises already $240$ million photometric measurements.  In order to
save computing time and  focus on objects that have not been classified
before, we limit ourselves to objects that are unlikely to be periodic by
choosing only objects with $Pf1 > 0.5$ from the catalogue in
\citet{2011A&A...529A..89D}. In order to include only objects that show
considerable variability we select objects with
$$std(F) > 3\langle \Delta F \rangle$$
where $F$ is the pre-search data conditioning simple aperture photometry
(PDCSAP, \citealp{2012PASP..124.1000S}) flux of the object and $\langle \Delta F
\rangle$ is the average photometric error. The light curves were then
preprocessed by taking the logarithm (base 10) of the flux and subtracting the
median of it. In order to get rid of individual spikes, caused by cosmic ray
hits, we go through the values $y_t$ of each light curve and replace them,
using a rather conservative constraint, according to:

$$
y_t = \left\{ \begin{array}{ll}
          y_t^{repl},\!\!\!\!\! & \mbox{if $(|y_{t-1} - y_t| > MAD)\mathrm{\land}(|y_{t+1} - y_t| >
MAD)$}\\
          y_t, & \mbox{otherwise} \end{array} \right.
$$
where $$y_t^{repl} = median ( \{y_{t+j} | j \in [-10,10]  \} )  $$ and 
$MAD$ is the median absolute deviation for the entire light curve $\bd{y}$.
On average, 150 of the 1624 observations have been replaced per time
series. After the  replacement of spikes, three further light
curves had to be excluded:
KIC4902072  appears to exceed the numerical range and therefore its
amplitude is occasionally swapping signs; KIC4346303 shows
dramatically higher amplitude and was removed in order to avoid introducing a
bias in the visualisation; KIC6117602 still shows some
heavy spikes after the spike removal which are probably of
an instrumental origin (damaged pixel) as well. Hence,
the final dataset was composed of 6,206 Kepler light curves.

\section{Results}
\label{sec:results}

\begin{figure*}
   \centering
   \begin{subfigure}[b]{0.7\textwidth}
       \includegraphics[width=\textwidth]{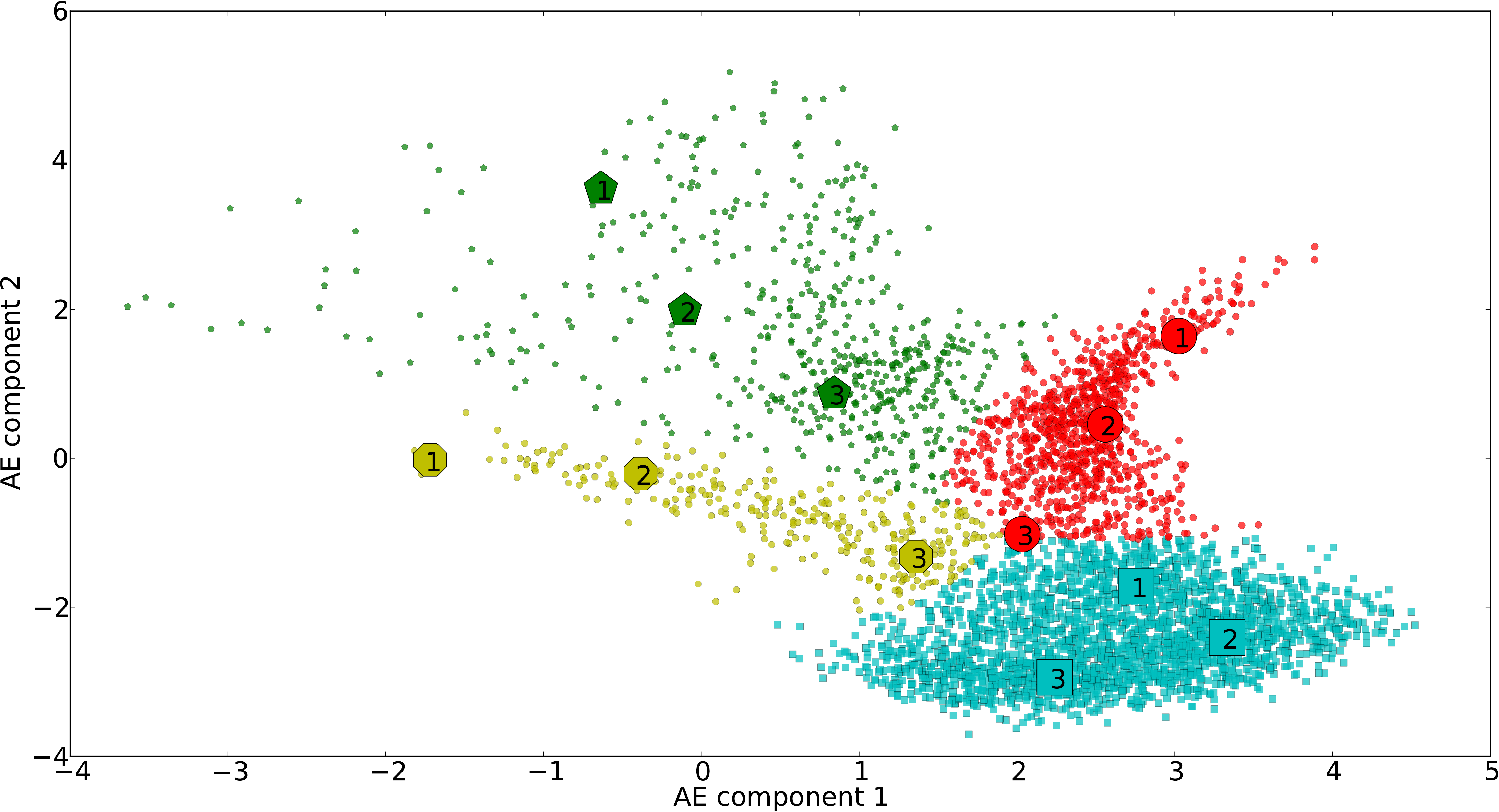}
       \caption{ESN-AE.\label{fig:cae}}
   \end{subfigure}
   \begin{subfigure}[b]{0.7\textwidth}
       \includegraphics[width=\textwidth]{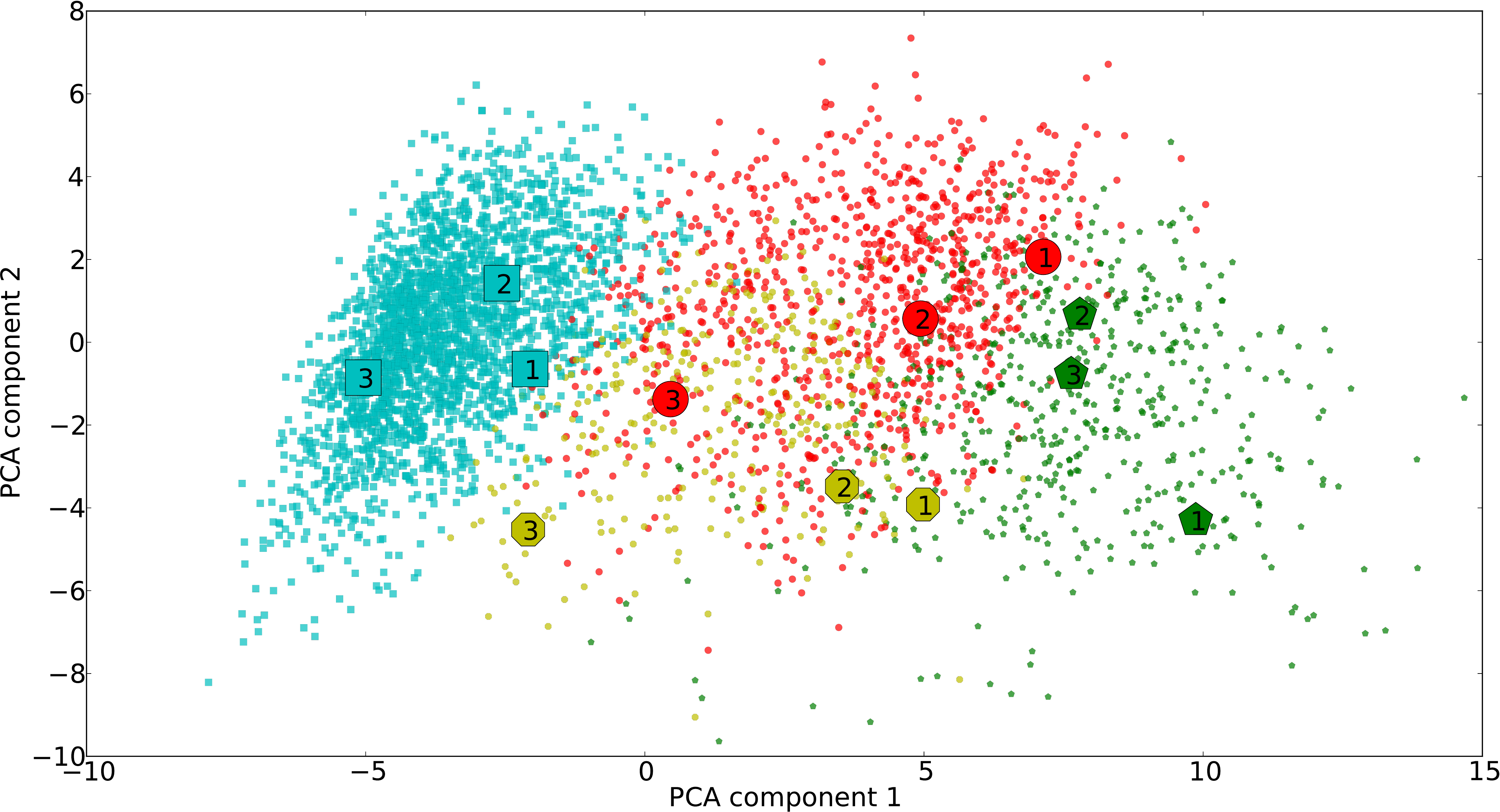}
       \caption{PCA on ESN weights. \label{fig:pca}}
   \end{subfigure}
   \begin{subfigure}[b]{0.7\textwidth}
       \includegraphics[width=\textwidth]{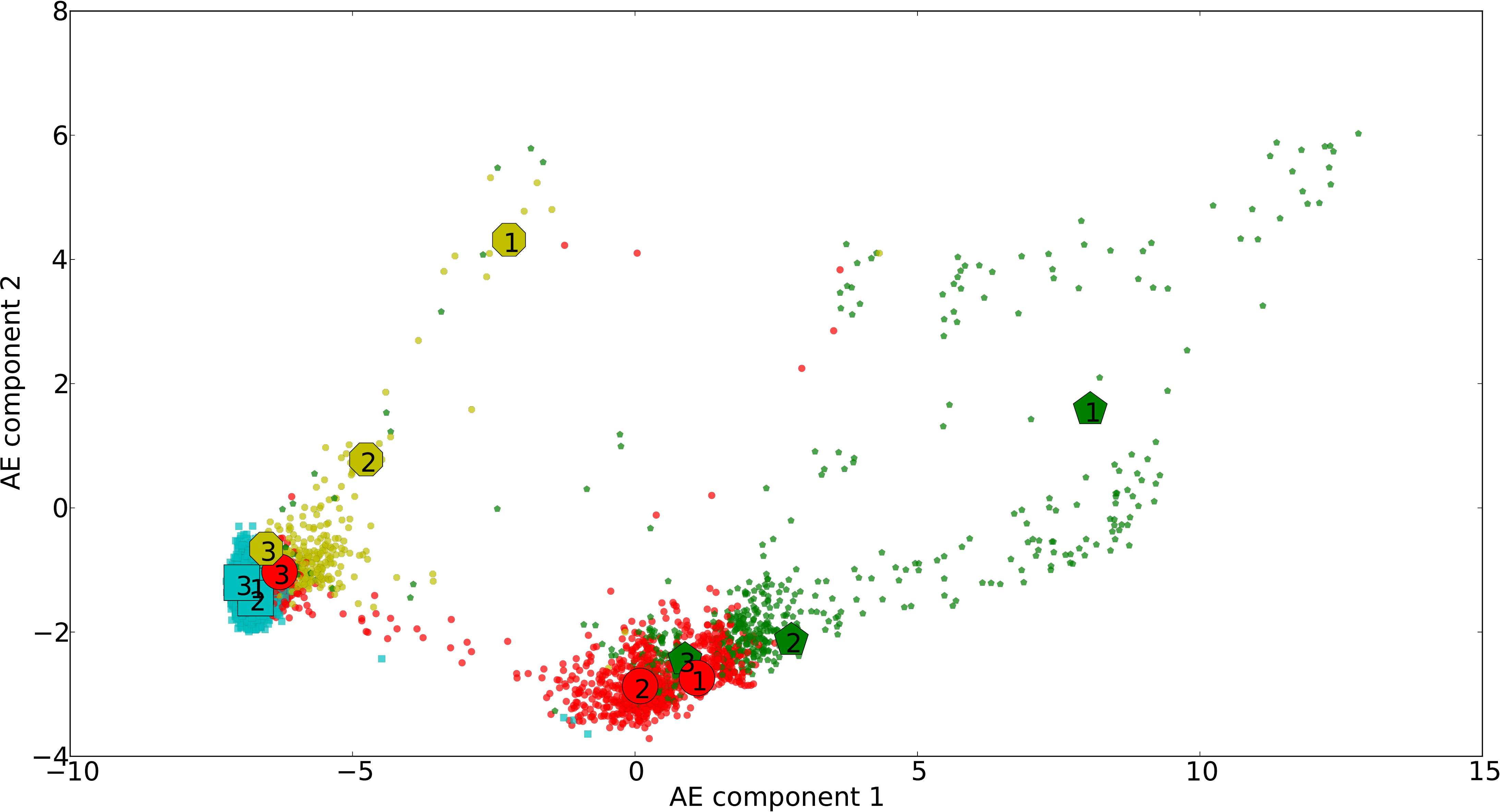}
       \caption{Plain autoencoder on ESN weights. \label{fig:pae}}
   \end{subfigure}
\caption{Visualisation of Kepler light curves using a coupled autoencoder
(a), a principal components analysis (b) and a plain autoencoder (c). The
visualisations are performed on the weights originating from the ESN. The
division into the classes is based on the ESN-AE visualisation. For
the sake of clarity, only 60\% of the points are plotted.
\label{fig:vis}}
\end{figure*}

In \mbox{Fig. \ref{fig:vis}}, we display the visualisations obtained by the
ESN-AE, PCA and a plain autoencoder.
We mention in passing that the parameter configuration used for the proposed method 
was an ESN with a hidden reservoir of $50$ neurons using the cycle architecture
\citep{Rodan2012} coupled to an autoencoder with a hidden layer of $10$
neurons. More information on the parameters, can be found in \citet{auto_enc}. 
For the plain autoencoder, we also used a hidden layer of $10$ neurons.

The visualisations in  \mbox{Figs. \ref{fig:pca}, \ref{fig:pae}}  are driven by
optimising the $L_2$ norm on the readout representations which means
that what they display is a two-dimensional projection of the readouts $\bd{w}$.
On the other hand, the ESN-AE in  \mbox{Fig. \ref{fig:cae}} is driven
by optimising  reconstruction on the sequences $\bd{y}$, and hence what we obtain
is a two-dimensional projection of the sequences. Though the readout parameters
$\bd{w}$ do capture important information about their respective sequences, we
do not expect in general that their visualisation will impart a meaningful
result. In other words, PCA and the plain autoencoder judge two light curves to
be similar, if their respective readouts are similar in the $L_2$ sense. The
ESN, however, judges two sequences to be similar if their corresponding
readouts result in similar sequences.

\begin{figure}
\includegraphics[width=0.47\textwidth]{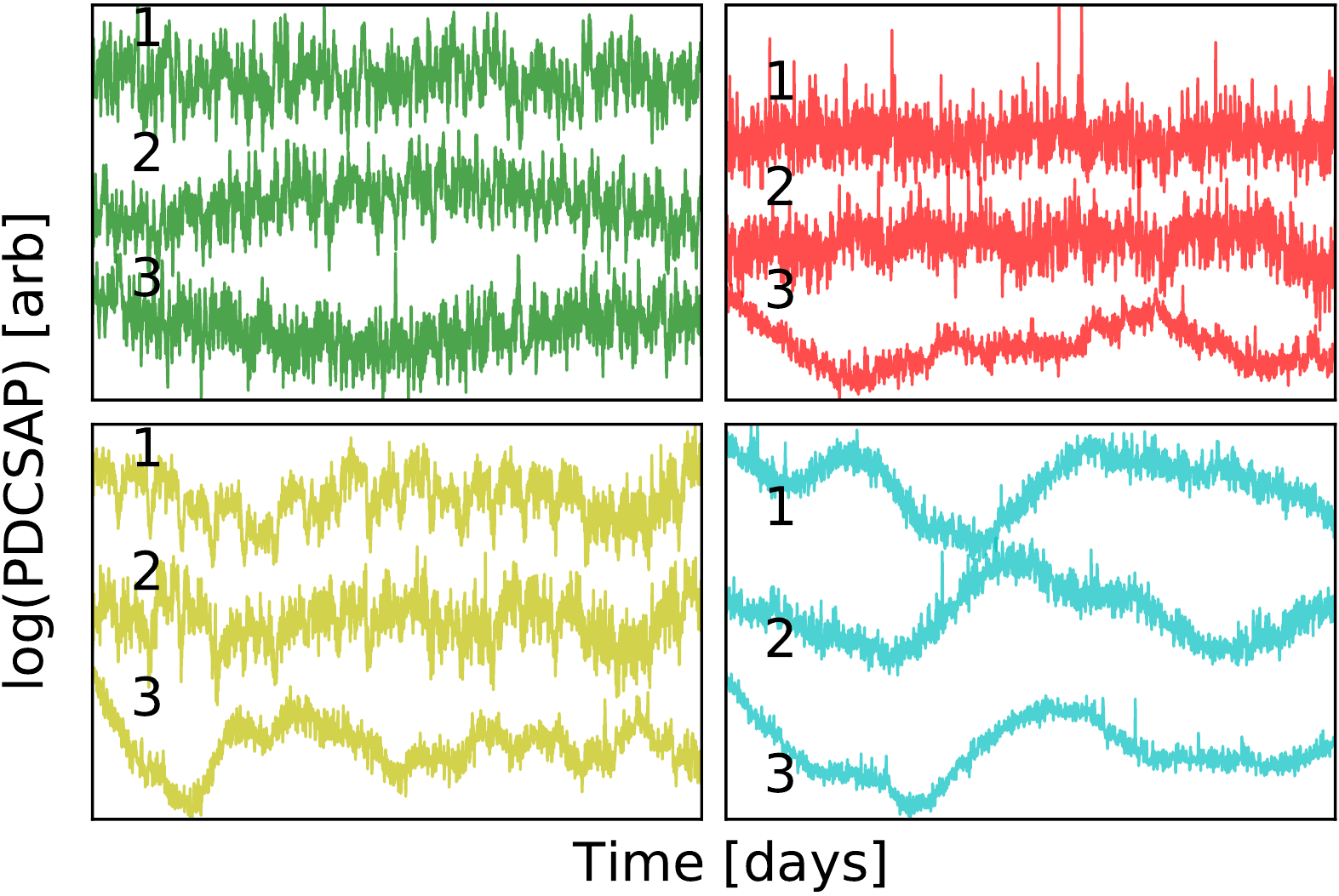}
\caption{Example light curves drawn from \mbox{Fig.
\ref{fig:vis}}. The x-axis corresponds to 33.16 days for each of the
images, ticks have been removed for clarity. The y-axis denotes the logarithm of
the PDCSAP flux with an arbitrary offset and scaling for better
visibility.\label{fig:samples}}
\end{figure}

We begin our analysis with the ESN-AE in \mbox{Fig.
\ref{fig:cae}}. After  inspecting this visualisation we noticed that there are, 
roughly speaking,  four regions of particular patterns which we  highlight with
four different colours. In every coloured region, we have marked three randomly
chosen light curves, which we display in \mbox{Fig. \ref{fig:samples}}.
It turns out that the four regions show distinct variability behaviour.
The yellow branch shows periodic, dip-like variability, fairly common
for rotating stars, while  the cyan curves show a prominent variability on time
scales of roughly ten days. The red curves are noisy  with small-amplitude and
short-term variability, while the green curves show a low-amplitude variability
as well but on longer time scales. 
We note that the red and green class are rather similar in their appearance.
We also show how the behaviour of sequences smoothly changes variability
regimes as we cross the borders of these four regions.  For instance, we note
how the yellow sequence \mbox{(1-2-3)} transits from short to long term
behaviour as we traverse from the edge of the yellow region towards the border
with the blue region. A similar behaviour is observed for the red sequences, as
we again traverse this region from sequence (1) towards sequence (3) close to
the blue and yellow border. This is of course not accidental, but rather by
construction, as the ESN-AE is placing similar dynamical behaviour in similar
locations.

We use the same colours in the other two visualisations, in order to show where
each coloured region in \mbox{Fig. \ref{fig:cae}} is mapped  in the PCA
visualisation of \mbox{Fig. \ref{fig:pca}} and in the plain-autoencoder
visualisation of \mbox{Fig.  \ref{fig:pae}}. We stress at this point that we are
aware that our interpretation of the map is subjective, and that other ways of
highlighting the visualisation are possible. However, compared to the other
visualisations, we find it easier in \mbox{Fig. \ref{fig:cae}} to discern some
structure. For instance, in \mbox{Fig. \ref{fig:pca}} we see that the readouts
$\bd{w}$ look all rather similar to PCA as no distinct structure stands out.
This could be perhaps attributed to the linear and inflexible nature of PCA, as
the plain-autoencoder in \mbox{Fig. \ref{fig:pae}} shows projections organised
in certain subgroups.

 With the selected example light curves at hand, we are able to
reinvestigate the meaning of the visualisations. The transition from short to
long term behaviour in the yellow sequence (1-2-3) to the cyan sequences is
fairly clear and apparent in all visualisations, even though, the distances are
judged quite differently between PCA, ESN-AE on one side and plain autoencoder
on the other. Strong support for the ESN-AE and the plain autoencoder comes from the
relative distance between red-3 and yellow-3 which is, compared to other
distances, quite large in the PCA visualisation. However, when inspecting the
light curves we note a strong similarity between these two and attribute the
high distance in the PCA to the inflexibility of the visualisation.
Two light curves of the green sequence (2,3) show  quite significant long-term
variability as opposed to the noise-dominated light curves in the red sequence
(1,2). In the plain autoencoder, the distances between those are judged
significantly different to the PCA and ESN-AE visualisations. While this alone
is not a strong argument, it is interesting to see that the noise-dominated
light curve from the green sequence (1) is projected very far away from the
noisy red ones, in terms of relative distances.

Additionally, the visualisation of the plain autoencoder shows a significant
overlap between the yellow, cyan and red light curves, even though, 
the variability behaviour of these three classes is
inherently quite distinct. We also note that the plain autoencoder, does not
enjoy the same smooth change in behaviour  observed in the ESN-AE visualisation.
This is very likely due to the missing link between reconstruction of weights
and reconstruction of sequences;  this hinders the plain autoencoder in
correctly interpreting the weights \bd{w} leading to a visualisation that
cannot be easily comprehended.

Given the above considerations and the principled objective function, we argue
that the ESN-AE model provides a better visualisation of the Kepler light curves. 
Of course, the evaluation of the visualisation is thereby not conclusive. 
In order to gain further insight as to how meaningful it is, 
we investigate in the next section how certain physical properties relate to the ESN-AE visualisation.



\begin{figure}
\centering
\includegraphics[width=.44\textwidth]{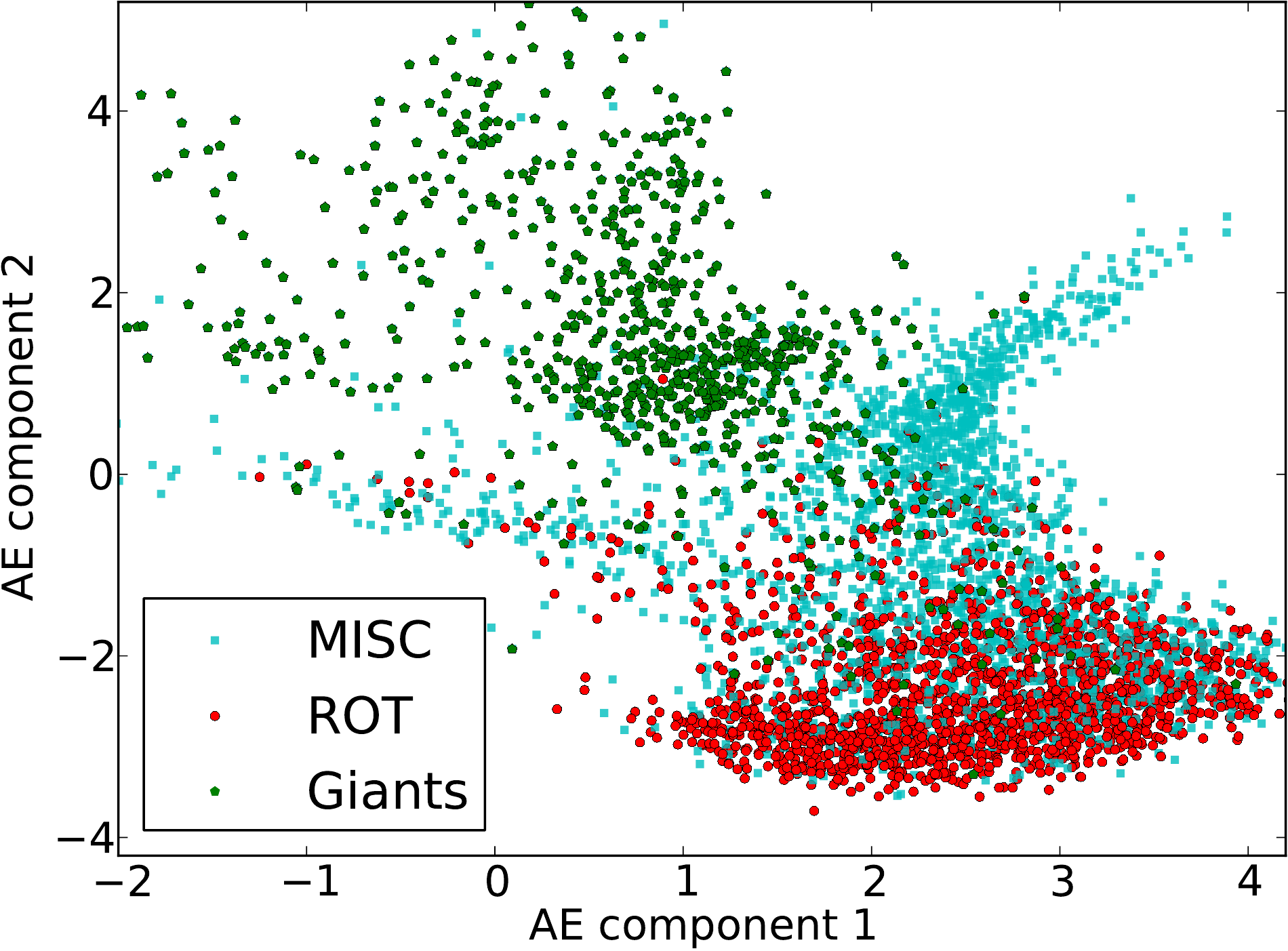}
\caption{Autoencoder visualisation with overplotted known
rotational stars from \citet{2014ApJS..211...24M}. Additionally, giant
stars as obtained by the temperature-$\log g$-criterion in \citet{2011MNRAS.414.2594H}
are shown. For the sake of clarity, only 60\% of the points are plotted.
\label{fig:rot}}
\end{figure}

\section{Discussion}
\label{sec:discussion}

\begin{figure*}
   \centering
   \begin{subfigure}[b]{.33\textwidth}
   	   \centering
       \includegraphics[width=.95\textwidth]{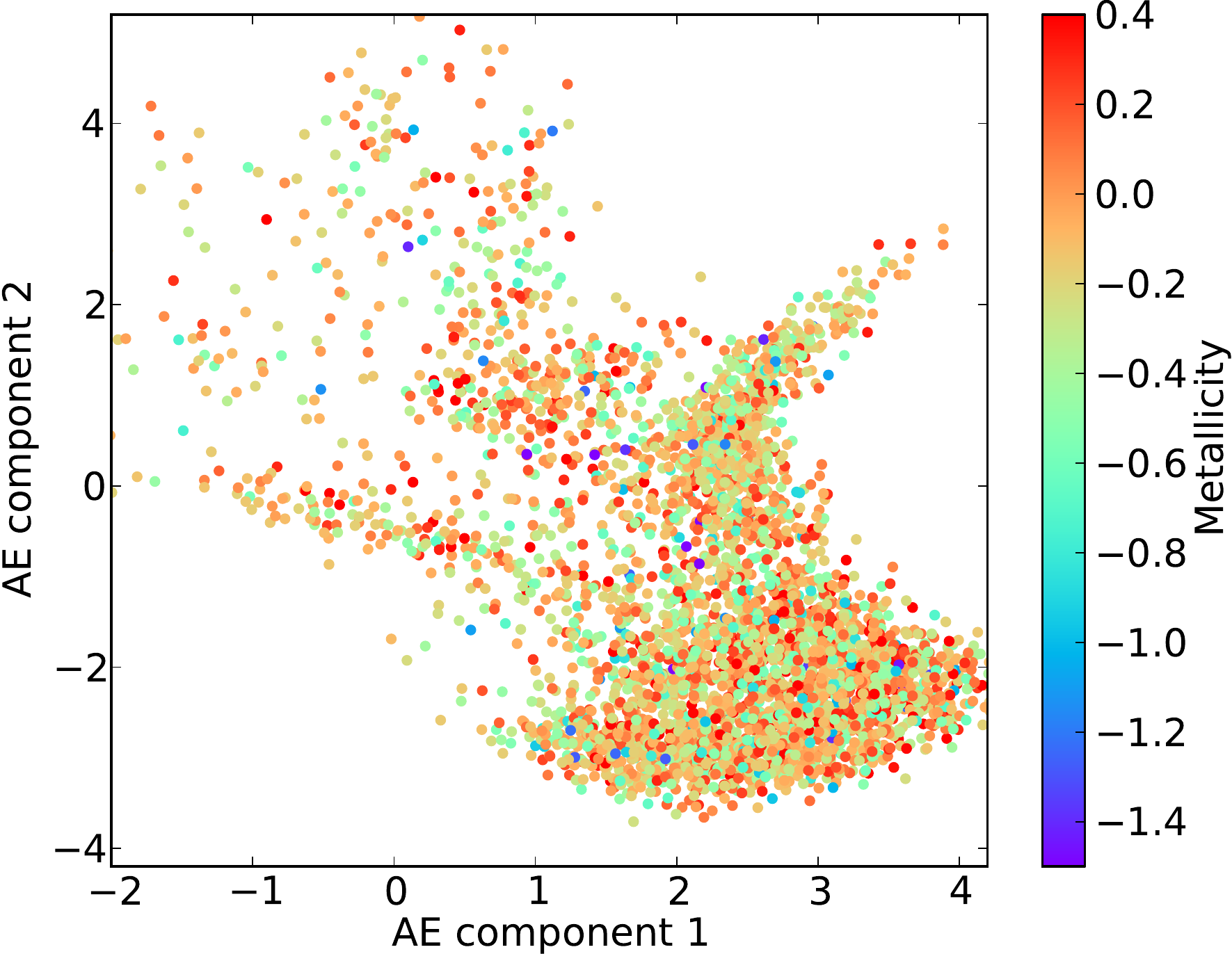}
		\caption{Metallicity\label{fig:metall}}
	\end{subfigure}
   \begin{subfigure}[b]{.33\textwidth}
       \centering
       \includegraphics[width=.95\textwidth]{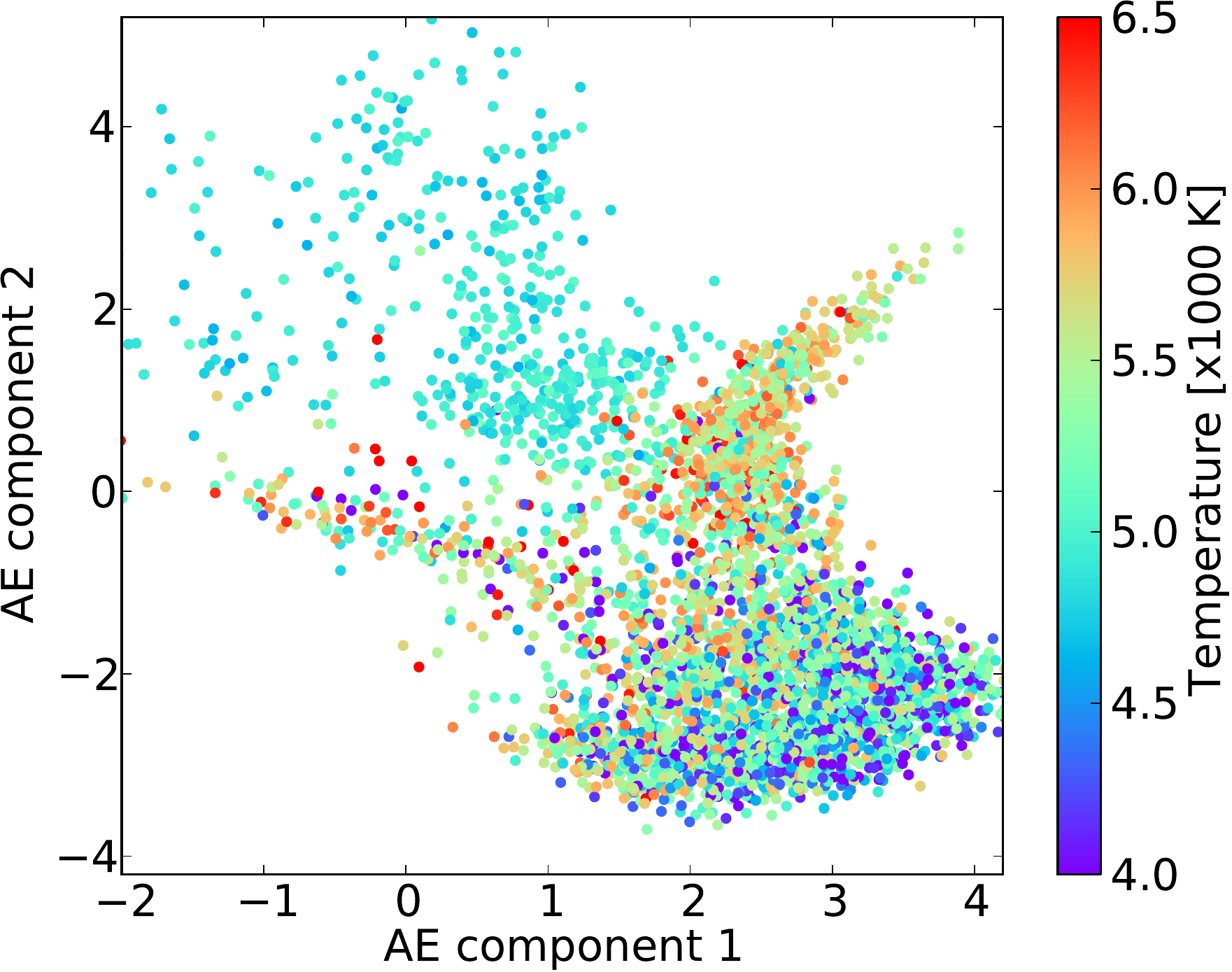}
		\caption{Temperature\label{fig:temp}}
	\end{subfigure}
   \begin{subfigure}[b]{.33\textwidth}
       \centering
       \includegraphics[width=.95\textwidth]{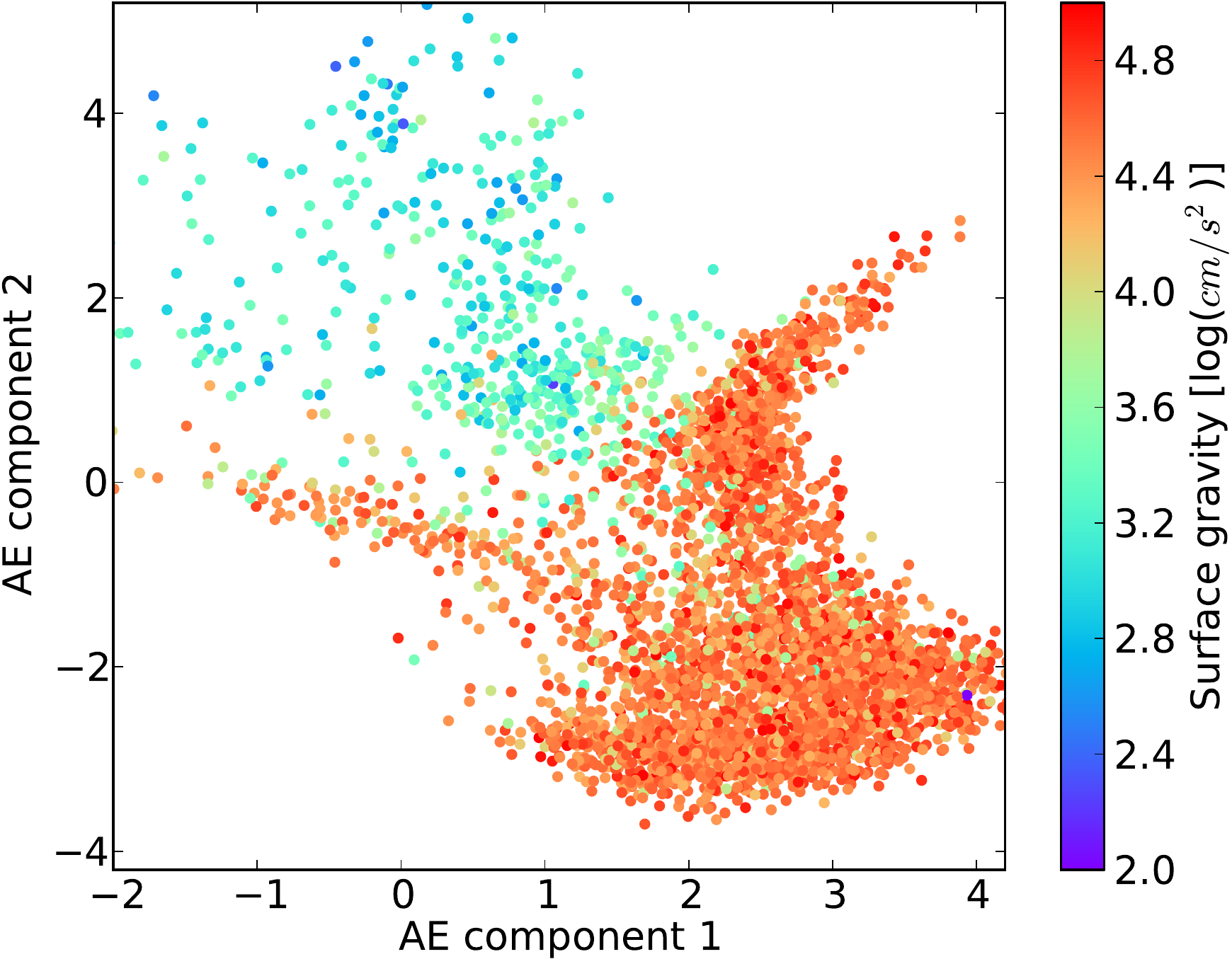}
       	\caption{Surface gravity\label{fig:surface}}
	\end{subfigure}
\caption{Visualisation of the coupled autoencoder with
overplotted properties from the KIC. A strong correlation
between the visualisation and the temperature as well as the
surface gravity is apparent. For the sake of clarity, only 60\% of the points
are plotted. Due to the presence of a few extreme values, we also limit
the colour scale to its present range to avoid colour saturation.
\label{fig:props}}
\end{figure*}

The central point of this work is the visualisation of time series with respect
to a given model, an ESN in this case. By design, the
ESN-AE delivers lower reconstruction errors on time series of all models
investigated in the previous section.
The other algorithms were only used in order to
highlight and discuss the impact of the coupling in the visualisation.
Therefore, the discussion focuses solely on the
visualisation results of this model.

The periodic variability with dip-like occurrences (yellow region) as well as
the long quasi-periodic behaviour of the cyan region are rather typical signs
of rotational stars. Fortunately, \citet{2014ApJS..211...24M} have scanned the
Kepler database for rotationally variable stars. In \mbox{Fig.
\ref{fig:rot}}, the location of the rotational stars found by them and being
part of our sample are highlighted in red. One can see that nearly exclusively
the cyan and the yellow regions are covered with rotationally variable stars,
indicating that many of the stars in these two regions are rotationally
variables as well.
Therefore, the objects in these classes present high-fidelity candidates
for rotational variability as well. It appears that especially the yellow
branch contains many formerly undetected objects which show similar dynamical
behaviour to the rotating stars. We note that such a clear dependence between
the visualisation and the location of the rotational stars is only visible in
the ESN-AE visualisation.

While the origin of most of the sources in the cyan and yellow region seems to
be resolved, the red and the green regions do not exhibit variability behaviour
which could uniquely identify their origin. Therefore, we inspected the data
provided by \cite{2011AJ....142..112B} in the Kepler Input Catalog 
(KIC\footnote{\href{http://adsabs.harvard.edu/abs/2009yCat.5133....0K}{http://adsabs.harvard.edu/abs/2009yCat.5133....0K}}). 
There, stellar model atmospheres \citep{2004astro.ph..5087C} were employed to 
photometric measurements of the Kepler sources in a probabilistic way. The
obtained temperature is uncertain to $\pm200$\,K and the estimate of the
logarithm of the surface gravity is uncertain to $0.4\,dex$.  With those
properties at hand, we highlight the giant stars in our sample, using the
temperature-$\log g$-criterion by \citet{2011MNRAS.414.2594H}. 5969 of the
visualised stars have a valid temperature and a valid $\log g$, of which 13\%
(777) are giants. The giant stars are overplotted in green in \mbox{Fig.
\ref{fig:rot}} as well. We note that they nearly exclusively populate the
green region and are well separated from the remaining main sequence stars.

In \mbox{Fig. \ref{fig:props}}, the individual properties from the KIC catalogue
are overplotted for the visualised light curves. It is apparent (\mbox{Fig.
\ref{fig:metall}}) that our visualisation 
does not correlate with metallicity. On the other hand, it has been noted in 
\cite{2011AJ....142..112B} that the uncertainty of this estimate is rather
high. In contrast to that, a clear correlation between our visualisation and the
temperature exists. Nearly all stars have temperatures between 4,000 and
6,000\,K \mbox{(F- and G-type)}, but in \mbox{Fig. \ref{fig:temp}} a clear
absence of hot or cold stars in the green region is apparent. Additionally, it
seems that the hot stars are mainly located in the red and yellow branch, while
cold stars can nearly exclusively be found in the bottom part of the cyan
region. It should be noted, that the uncertainty in the temperature estimation
is not sufficient to explain this separate behaviour. Further support for the
visualisation comes from the distribution of surface gravities in \mbox{Fig.
\ref{fig:surface}} which reflects that nearly all giants are located in the
green region. The correlation between variability behaviour and stellar
properties, such as the surface gravity, has also been studied in
\citet{2013Natur.500..427B}. There, a correlation between the 
variability - that remains after subtracting the average brightness in 8 hour bins - and the
surface gravity was discovered. In \citet{2010ApJ...723.1607H}, red giant
variability was detected by investigating the correlation between frequency
separation and the frequency of maximum power. In both publications,
the goal was to detect giant stars based on their variability behaviour.
To that end, explicit physical knowledge was invoked.
It is therefore very pleasing to see that the proposed unsupervised method can
also separate the giants based on their variability alone, without 
any knowledge about periodicity or other physical properties.

So far, this work considered only the unsupervised task of dimensionality
reduction of astronomical time series. However, it is possible to extend it to
supervised tasks such as classification or regression.
We briefly make an example on classifying the giants previously identified  in
the visualisation. We trained two classifiers. The first one is a random
forest\footnote{We use a $1,000$ trees with balanced class weights and a
3-fold cross-validation scheme.} trained on all $29$ features in the catalogue
by \citet{2011A&A...529A..89D} yielding an accuracy of  $96.0\pm 0.4\%$.
The second one is also a random forest trained on the two visualisation
components as inputs yielding a higher classification accuracy of
$96.6\pm 0.2\%$. While we emphasise that this is not a conclusive statement on its own,
it is remarkable to see that the classification on just the two visualised
components performs as well as the one on the $29$ (physically motivated) features.
The two visualisation components describe similarity in terms of the dynamical
behaviour and appear to be highly informative. In addition, one can
consider utilising the ESN readout weights $\bd{w}$ as features. These can then
be used as inputs to a classifier with an objective (cross entropy) function
that quantifies classification accuracy rather than reconstruction error.




\section{Conclusions}
\label{sec:conclusions}

In this work, a new approach to visualise regularly sampled time series was
presented. As opposed to  visualisation algorithms in astronomy,
the presented one does not require any pre-alignment of the data and respects
the sequential nature of the time series. Besides that, it is capable to deliver
a shift invariant vector representation for sequential data of variable length.
Compared to the common use of visualisation algorithms in astronomy, we do not
employ the dimensionality reduction directly on the data, but on model
parameters instead. We strongly advocate the use of sequential models to
describe time series in astronomy and highlight the advantages of those using
an ESN.
The proposed ESN model returns a fixed-length vector representation for a given
sequence. This in turn can then be fed to a visualisation algorithm. In order to
enhance the meaning of the visualisation, we measure the reconstruction error
not in terms of reconstructing the model parameters but by measuring the direct
implications on the reconstruction of the original light curves. This approach
provides a powerful objective function which also leads to a more meaningful
visualisation. 

The proposed visualisation was demonstrated on a selected
subset of $6,206$ light curves of variable stars. We studied the quality
of the plain and ESN-AE algorithms empirically and concluded that
the proposed coupled visualisation algorithm returns results that are easier
to interpret.
Further support for the proposed visualisation comes from the physical
properties of the stars that have been derived from the time series data. With
those, we can clearly see that the green cluster is mainly made up from
giants with surface temperatures of $5,000-5,500$\,K and a significantly
lower surface gravity than the main sequence stars in the sample.
Besides that, it appears that also separate regions are populated by hot
($>5,500$\,K) and cold stars ($<5,500$\,K). 
It is interesting that these physical properties (surface gravity, temperature)
show up in the visualisation the way they do, as the underlying model is not aware of them.
The correlation between physical properties and variability has been identified
in other works by explicitly looking for it using tailored features. The proposed
visualisation confirms this correlation thus showing that these physical properties
are inherent in the light curve dynamics.
We speculate that the ESN readout representation could be further used 
in regression (e.g. predict surface gravity)
and classification (e.g. main sequence versus giant stars) tasks.


The presented approach is modular in the sense that
parts of it can be simply replaced. The autoencoder is merely a convenient
candidate but other visualisation algorithms could be used instead, perhaps
with more favourable computational properties. Additionally, the underlying
dynamical ESN model,  could be replaced by other models capable of describing
sequential data, such as auto-regressive models (ARMA). Finally, the approach
is not limited to sequences and in principle other types of astronomical data,
given a suitable model, can be visualised in the same fashion. Currently, the
visualisation of SDSS spectra using a blackbody model is investigated.

\label{lastpage}
\bibliographystyle{mn2e}
\bibliography{bibliography.bib}
\end{document}